\documentclass[twocolumn,prb,10pt,showpacs,preprintnumbers,superscriptaddress,amsmath,amssymb,citeautoscript,aps]{revtex4-1}
\usepackage{graphicx}
\usepackage{dcolumn}
\usepackage{subfigure}
\usepackage{color}
\usepackage{bm}
\usepackage[hidelinks]{hyperref}
\hypersetup{
    colorlinks,
    citecolor=blue,
    filecolor=blue,
    linkcolor=blue,
    urlcolor=blue
}

\newcommand{\im}{\textrm{i}}

\DeclareMathOperator{\e}{e}
\DeclareMathOperator{\C}{\mathcal{C}}

\begin{document}

\title{Signature of a topological phase transition in long SN junctions in the spin-polarized density of states}
\author{M. Guigou}
\email{marine.guigou@gmail.com}
\affiliation{Institut de Physique Th\'eorique, CEA/Saclay, Orme des Merisiers, 91190 Gif-sur-Yvette Cedex, France}
\author{N. Sedlmayr}
\affiliation{Institut de Physique Th\'eorique, CEA/Saclay, Orme des Merisiers, 91190 Gif-sur-Yvette Cedex, France}
\affiliation{Department of Physics and Astronomy, Michigan State University, East Lansing, Michigan 48824, USA}
\author{J. M. Aguiar-Hualde}
\affiliation{Institut de Physique Th\'eorique, CEA/Saclay, Orme des Merisiers, 91190 Gif-sur-Yvette Cedex, France}
\author{C. Bena}
\affiliation{Institut de Physique Th\'eorique, CEA/Saclay,
Orme des Merisiers, 91190 Gif-sur-Yvette Cedex, France}
\affiliation{Laboratoire de Physique des Solides, UMR 8502, B\^at. 510, 91405 Orsay Cedex, France}

\date{\today}

\begin{abstract}
We investigate the spin texture of Andreev bound states and Majorana states in long SN junctions. We show that measuring the spin-polarized density of states (SPDOS) allows one to identify the topological transition. In particular, we find that its total component parallel to the wire is non-zero in the topological phase for the lowest-energy state, while vanishing in the trivial one. Also, the component parallel to the Zeeman field is symmetric between positive and negative energies in the topological phase and asymmetric in the trivial phase. Moreover the SPDOS exhibits a moderate accumulation close to the SN boundary which changes sign when crossing the topological transition. We propose that these signatures may allow one to unambiguously test the formation of a topological phase via spin-resolved transport and STM measurements. 
\end{abstract}


\maketitle

\section{Introduction} Recently the problem of the formation and detection of Majorana fermions has drawn a lot of attention.\cite{Nilsson2008,Akhmerov2009,Law2009,Fu2009,Alicea2012} The formation of Majorana states has been predicted by many theoretical works \cite{Moore1991,Read2000,Kitaev2001,Stern2004,Fu2008,Sato2009b,Sau2010,Potter2011,Cook2011,Wong2013,Sedlmayr2015,Sedlmayr2015a,Sedlmayr2015b,Sedlmayr2016}. One of the systems predicted to exhibit Majorana fermions is a one-dimensional semiconducting wire with strong spin-orbit coupling such as InSb\cite{Mourik2012,Deng2012} or InAs\cite{Das2012}, in the proximity of an superconducting substrate, and in the presence of a Zeeman magnetic field \cite{Lutchyn2010,Oreg2010}. However, an unambiguous experimental detection of Majorana fermions remains unattained despite many promising experiments \cite{Mourik2012,Deng2012,Das2012}. 

In the quest to find an unequivocal fingerprint of the topological superconducting phase, which can support Majorana states, we study the spin texture of the Andreev bound states (ABS) and of the Majorana bound states (MBS) in superconductor-normal (SN) junctions. 
As it has been shown in Refs.~\onlinecite{Black-Schaffer2011} and \onlinecite{Chevallier2012}, in such junctions extended Majorana states can form inside the normal link  in the topological phase, which may lead to intriguing phenomena such as fractional Josephson effects\cite{Fu2009,Kitaev2001,Lutchyn2010,Law2011,Asano2013}. The dependence of the MBS and ABS on the system parameters has been explored quite thoroughly in the past\cite{Nilsson2008,Akhmerov2009,Fu2009,Kitaev2001,Black-Schaffer2011,Chevallier2012,Tanaka2009,Linder2010,Tanaka2010,Tanaka2012}, and the spin-polarized transport through these states has been touched upon\cite{Liu2013,Lee2014,He2014}, but the spatial spin texture of these states in a topological system has until now been generally overlooked\cite{Nagai2014}. 

We consider quasi-1D wires with both longitudinal and transverse Rashba spin-orbit coupling, in the presence of a Zeeman field, and  for which a section of the wire was brought in the proximity of a superconductor to make a topological SN junction. We find that the spin texture for such systems exhibits characteristic features when crossing the transition between trivial and topological phases which could be used to identify this transition using techniques such as spin-resolved STM and transport measurements. 
In particular we find that the spin-polarized density of states (SPDOS) of the MBS integrated for the entire SN junction, unlike that of the ABS in the topologically trivial phase, is nonzero for the spin component parallel to the wire.  Thus we prove that measuring the total SPDOS of the lowest energy state, (via current injection from spin-polarized leads for example), provides one with a good estimate for the topological invariant, distinguishing between the topologically trivial and non-trivial phases. We propose this as a distinguishing feature of the phase transition.
\begin{figure}
\includegraphics[width=0.95\columnwidth]{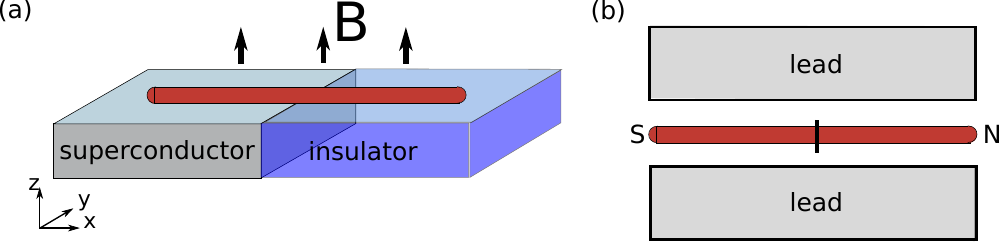}
\caption{(Color online) Schematics of the system under consideration. (a) A semiconducting wire with strong spin-orbit coupling is brought in proximity with a SC substrate over one extended region, to make a long SN junction. Panel (b) shows the view from above of the coupling to two leads for the spin resolved transport experiment. The two leads are both weakly coupled to the entire wire, the black line shows the boundary between the normal and superconducting regions of the wire. Between these layers should be an insulating layer to screen the bulk superconductor from the leads and to prevent any tunnleing between them.}
\label{Fig1}
\end{figure}

Furthermore in the strictly 1D limit when only a single band is occupied
the SPDOS of the MBS and ABS exhibits a moderate accumulation close to the SN boundary.  When one crosses into the trivial phase, the two zero-energy MBS split into two regular ABS whose spin polarization shows qualitatively different features than in the topological phase. In particular the asymmetry between the state with positive energy and its negative-energy counterpart becomes very pronounced. The most striking feature is a change of sign for the spin polarization close to the SN boundary between the topological and the trivial phase.

\section{Model} We consider a generic SN junction (see Fig.~\ref{Fig1}) described by the following Bogoliubov-de Gennes tight-binding Hamiltonian for a wire of width $N_y$:
\begin{eqnarray}
\label{HamiltonianSNS}
H&=&\sum_{\ell=1}^{N_y}\sum_{j=1}^N\Psi^\dagger_{j\ell}\left[(t-\mu){\bm\tau}^z+B{\bm\sigma}^z-\Delta_j{\bm\tau}_x\right]\Psi_{j\ell}\nonumber\\&&
-\frac{1}{2}\sum_{\ell=1}^{N_y-1}\sum_{j=1}^N\left[\Psi^\dagger_{j\ell}(t-\im\alpha{\bm\sigma}^x){\bm\tau}_z\Psi_{j,\ell+1}+\textrm{H.c.}\right]\\&&
-\frac{1}{2}\sum_{\ell=1}^{N_y}\sum_{j=1}^{N-1}\left[\Psi^\dagger_{j\ell}(t+\im\alpha{\bm\sigma}^y){\bm\tau}_z\Psi_{j+1,\ell}+\textrm{H.c.}\right]\,,\nonumber
\end{eqnarray}
written in the Nambu basis $\Psi_{j\ell}=(\psi_{\uparrow,j\ell},\psi_{\downarrow,j\ell},\psi^{\dag}_{\downarrow,j\ell},-\psi^{\dag}_{\uparrow,j\ell})^T$,
where $t$ is the hopping amplitude,  $\mu$ is the chemical potential, $\alpha$ is the spin-orbit coupling, $\Delta_j$ is the on-site pairing potential and $B$ is the Zeeman field. Here ${\bm\sigma}$ and ${\bm\tau}$ are the Pauli matrices acting respectively in the spin and the particle-hole subspaces, and the operator $\psi^\dag_{\sigma ,j\ell}$ creates a particle of spin $\sigma=\uparrow,\downarrow$ at site $(j,\ell)$ in the quasi-1D wire lattice.
For an SN junction of total length $N=L_1+L_2$ (see Fig.~\ref{Fig1}), we have
\begin{equation}\label{delta}
\Delta_j=\left\{\begin{array}{ll}\Delta\e^{\im\phi_1}&\textrm{ if }j\leq L_1\textrm{ and}\\
0&\textrm{ if }L_1<j\,.
\end{array}\right.
\end{equation}

Exact diagonalization of this single-particle Hamiltonian directly gives us access to the eigenvalues $\varepsilon_n$ with $n=1,2,\ldots d$, $d=4NN_y$ and the eigenvectors which we write in the Nambu basis
as $(u^n_{\uparrow,j\ell},u^n_{\downarrow,j\ell},v^n_{\downarrow,j\ell},v^n_{\uparrow,j\ell})$.  Throughout the paper we will consider $t=\hbar=1$, so that all energies are expressed in units of $t$.
The local density of states (LDOS) and the three local SPDOS components are defined similarly to Ref.~\onlinecite{Sticlet2012}:
\begin{eqnarray}
\label{dos}
\rho_{j\ell}(\omega)/S^z_{j\ell}(\omega)&=&\sum_{n=1}^{d} \nu(\omega-\varepsilon_n) \left(|u_{\uparrow,j\ell}^{n}|^2\pm |u_{\downarrow,j\ell}^{n}|^2\right)\,,\\\nonumber
S^{x/y}_{j\ell}(\omega)&=&\sum_{n=1}^{d} \nu(\omega-\varepsilon_n)2 {\rm Re/Im} \left(u_{\uparrow,j\ell}^{n*} u_{\downarrow,j\ell}^{n}\right)\,,\\\nonumber
\end{eqnarray} 
where $\nu(\omega)=e^{-\frac{\omega^2}{\gamma^2}}/(\sqrt{\pi}\gamma)$ is a Gaussian peak, with a width $\gamma=0.002t$. This broadening allows us to accurately take into account the overlapping contributions from different bands when they get close to each other or cross. Finally the total SPDOS components along the direction $\hat{n}=\hat x,\hat y,\hat z$ for a state $m$ are defined as 
\begin{equation}
S^{\hat{n}}=\sum_{j\ell}S^{\hat{n}}_{j\ell}(\omega).
\end{equation}

\section{Quasi 1D wires}

\begin{figure}
\includegraphics*[width=0.95\columnwidth]{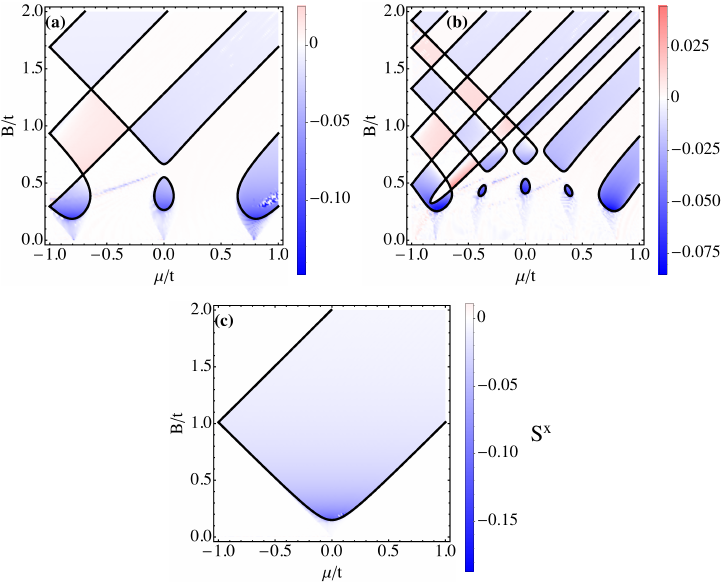}
\caption{(Color online) The total spin polarized DOS $S^x$ for the lowest energy state as a function of magnetic field and chemical potential. The phase diagram obtained using topological invariant calculations is very accurately recovered: the blue and pink regions correspond to a non-zero total SPDOS while the solid lines correspond to the analytically calculated phase boundaries.\cite{Sedlmayr2015b,Sedlmayr2016} We use $\alpha=0.5t$ and $\Delta=0.15t$ in all panels. The widths of the wires are (a) $N_y=3$, (b) $N_y=7$, and (c) the strictly-1D limit with $N_y=1$. We take $N=301$.}
\label{Fig2}
\end{figure}

We analyze the low-energy states numerically, using the MathQ code\footnote{See http://www.icmm.csic.es/sanjose/MathQ/MathQ.html}. As examples we consider $N_y=1$, $3$, and $7$. These systems can be brought into topologically non-trivial phases  by changing for example the Zeeman field and the chemical potential. The lowest-energy state in the system, which in the topological phase is a MBS, shows a clear signature of the non-trivial topology in its total $S^x$ component of the SPDOS (integrated over the entire SN system). In particular in the topological region the x-component of the total SPDOS is non-zero, while in a trivial phase this vanishes. We illustrate this in Fig.~\ref{Fig2} where we plot the spatially integrated SPDOS $S^x$ at the energy corresponding to the lowest state in the system, as a function of magnetic field and chemical potential. We see that the regions in which the total SPDOS is non-zero (corresponding to the regions colored in blue and pink respectively) overlap exactly with the topological phases predicted for these systems in previous work \cite{Sedlmayr2015b,Sedlmayr2016}. The limits of the topological phases calculated in these works using the topological invariant is denoted by the black lines in Fig.~\ref{Fig2}.  

This is a very interesting observation, as the totalSPDOS can be measured using spin-polarized transport (tunneling injection from spin-polarized leads, quantum dots, etc.). In such an experiment, if the spin of the leads is parallel to the axis of the wire, the value of the conductance will show abrupt changes at the boundaries between the topological and trivial phases, thus allowing one to test the existence of these boundaries, and thus the formation of a topological phase and MBS.

We can try to understand intuitively the origin of this phenomenon. A MBS in an SN junction is localized close to the end of the S region \cite{Chevallier2012}. Its counterpart is extended throughout the normal region, much like an ABS\cite{Chevallier2012}, and, for a large enough superconducting gap and a long enough normal region, it oscillates more or less uniformly. Thus the total contribution of the MBS to the $S^x$ coming from the normal region is close to zero. However, in the superconducting region, there is a strong anisotropy and most of the DOS is localized at the external end, where the Majorana state is formed. The Majorana state has a non-zero component of the $S^x$ which thus contributes to a total non-zero SPDOS for the lowest-energy state of the junction. Note that all the ABS states, which are uniformly extended throughout the normal region, have $S^x\approx0$, thus the measure of the SPDOS is not very sensitive to the accuracy of the measurable energy width, though one should always perform the measurements at the energy corresponding to the lowest energy state. 

\section{1D limit} 
\begin{figure}[h!]
\includegraphics[width=7.5cm]{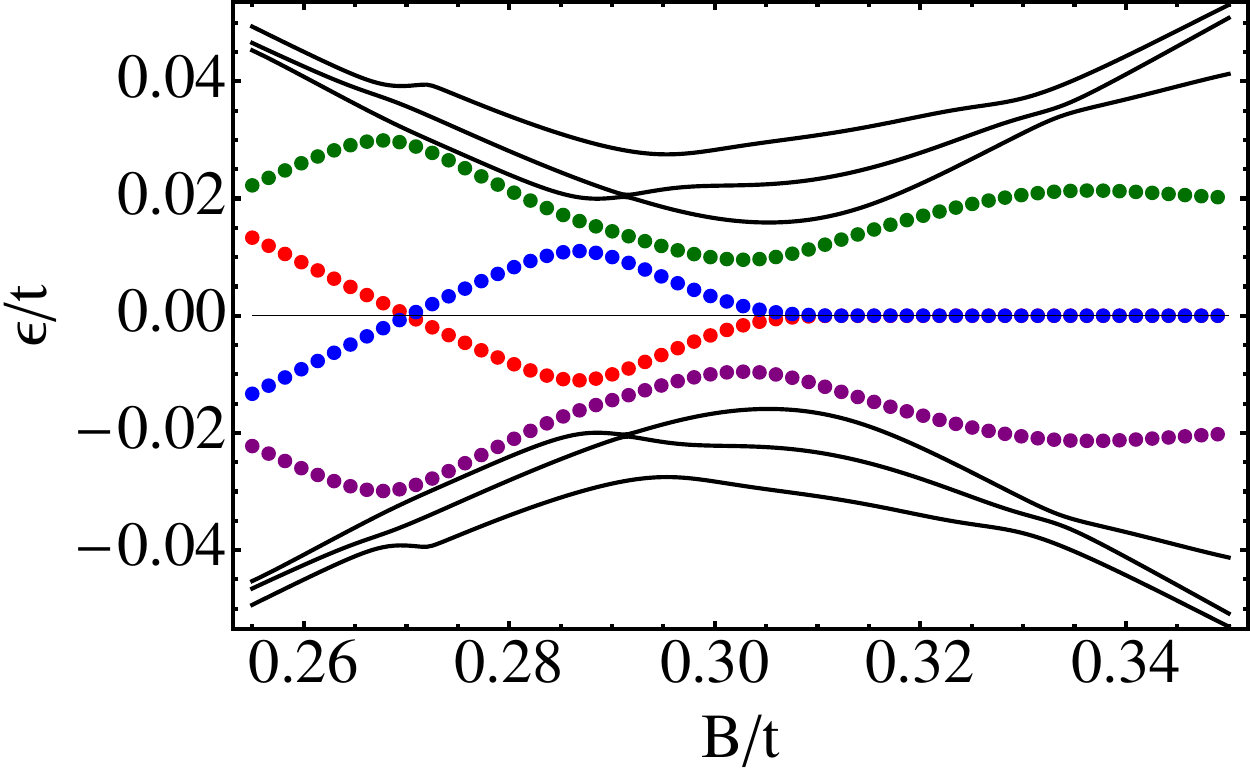}
\caption{(Color online) The low-energy states as a function of $B$ in an SN junction. We set $\Delta=0.3t$, and $\alpha=0.15t$. }
\label{Fig3}
\end{figure}

Moreover, in the 1D limit there are additional signatures of the topology in the LDOS and local SPDOS. 
We consider a 1D SN junction (hence $N_y=1$), with $L_1=L_2=50$ sites, $\Delta=0.3t$, and $\alpha=0.15t$. We focus on $\mu=0$, i.e.~the bottom of the band. In Fig.~\ref{Fig3} we plot the dependence of the energy levels near zero energy as a function of $B$. We note that the transition from the trivial to the topological phase is marked by a merging of the two lowest-energy ABS into MBS. In the thermodynamic limit this transition will occur at the critical field $|B^*|=|\Delta|$.\cite{Lutchyn2010,Oreg2010} \footnote{However the critical field is shifted by $\approx 1/L_1$ in a finite-size system, and by inspection of Fig.~\ref{Fig3} for our system this occurs at a critical value of $B^*\approx\Delta+0.5/L_1\approx0.31$.}
  
\begin{figure}
\includegraphics*[width=0.95\columnwidth]{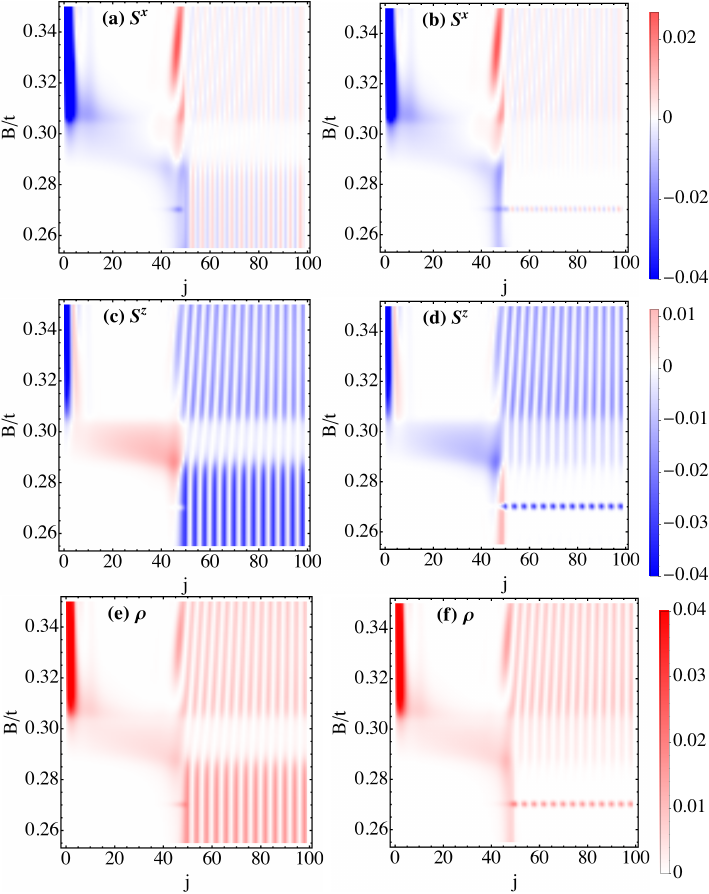}
\caption{(Color online) The $S^x$ and $S^z$ spin polarization components, and the LDOS, as a function of position and $B$ when following the evolution of the MBS into an ABS doublet. For the first ABS we follow the red line in  Fig.~\ref{Fig3}  (panels a,c,e) and for the second one the blue line in Fig.~\ref{Fig3} (panels b,d,f). The parameters are $\Delta=0.3$, and $\alpha=0.15$. To enhance contrast, the large external MBS peaks are `cut'-out, so that their actual intensity does not show on the color scales. Their respective intensities are $-0.145$ in a) and b) panels, $-0.0926$ in c) and d) panels and $0.086$ in e) and f) panels.}
\label{Fig4}
\end{figure}

In Fig.~\ref{Fig4} we plot the two non-zero spin components $S^x$ and $S^z$, as well as the LDOS, as a function of position and Zeeman field, at energies given by those of the lowest energy eigenstates  (i.e following the red and respectively blue lines in Fig.~\ref{Fig3}).  For the system considered here $S^y=0$. In the topological phase the LDOS and $S^z$ exhibit a peak at the external edge of the superconducting section, corresponding to a localized MBS. In the normal region both the MBS and the ABS show a roughly-uniformly-distributed weight in LDOS and $S^z$.  Moreover $S^x$ exhibits two peaks, a large one at the external end of the superconductor (SC), and a smaller one of opposite sign at the SN interface. In the trivial phase we note that the LDOS and both spin components show a very marked accumulation close to the SN boundary for the positive energy state, and approximately uniform amplitude oscillations with a small $S^x$ accumulation at the boundary for the negative energy state. 
When crossing the phase transition, the Majorana state localized at the external end of the SC disappears, as expected. 

The first key observation to be made is that the $S^x$ accumulation at the SN interface is changing sign when the systems goes from the trivial to the topological phase. 
Thus, measuring  the $S^x$ dependence on position and Zeeman field would allow one to detect the topological transition by an observation of a sign change in $S^x$ at the interface. This feature is unique to the SPDOS, and cannot be observed in the non-spin-resolved DOS. 

The second observation is that the first pair of non-zero energy states, (i.e.~those with positive and negative energies of equal magnitude), exhibit identical LDOS, as well as identical $S^x$ and $S^z$ components in the topological phase, but different SPDOS and LDOS in the trivial phase.  Thus measuring for example $S_z$ can provide one with another tool to identify the transition between the topological and trivial state via a measure of the asymmetry of the SPDOS between positive and negative energies. As mentioned before this could be used as a signature of the transition between the topological and trivial phase visible both in spin-resolved transport, as well as in spin-polarized STM measurements.

Note that these observations are characteristic to the strictly 1D wire and cannot be recovered in the quasi-1D wires with transverse Rashba.

\begin{figure}
\includegraphics[width=0.8\columnwidth]{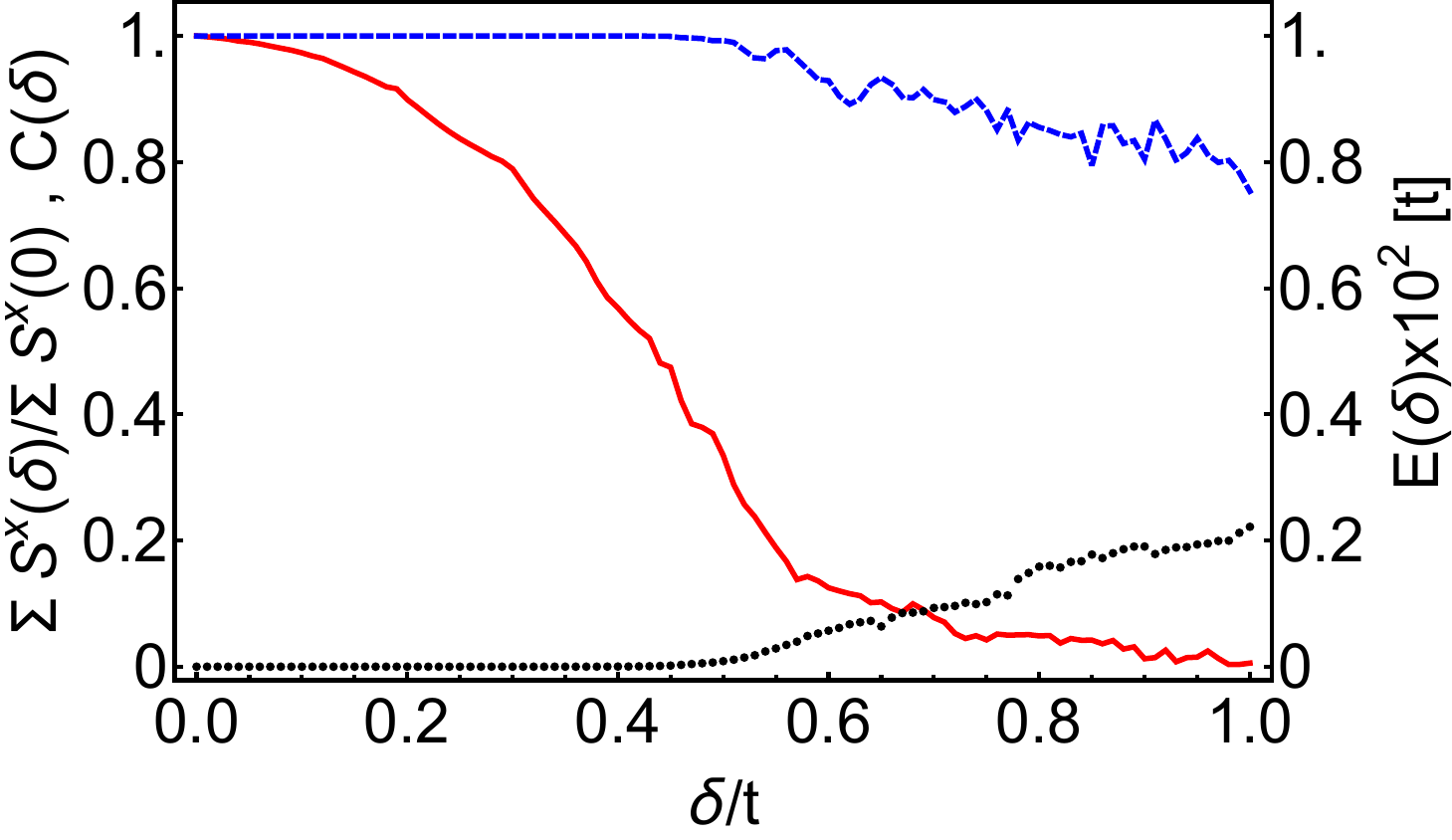}
\caption{(Color online) The total  $S^x$ spin polarization (red, solid) as a function of disorder strength, compared to the Majorana polarization $C(\delta)$ (blue, dashed) and the energy of the lowest energy state $E(\delta)$ (black, dotted), at a magnetic field strength of $B=0.35t$, all other parameters as for Fig.~\ref{Fig4}. The total spin has been scaled compared to its value in the clean limit $\sum_jS^x(\delta=0)=-0.190$. To obtain the disorder average a random potential varying between $-\delta\to\delta$ was assigned to each site and 100 different realizations of disorder were averaged over.}
\label{Fig5}
\end{figure}

\section{Universality of our results}
The observations described here are quite general as we have shown by taking different numbers of coupled wires for the quasi-1D systems. We have checked that they are valid as long as the length of the normal section is long enough (larger than the superconducting penetration depth, coherence length and spin-orbit length). Also we have checked that the results are qualitatively unchanged by scattering at the SN boundary for not too large interface scattering values (See the appendices for more details), as well as for weak static disorder. As detailed in Fig.~\ref{Fig5}, for randomly distributed impurities with strength varying between $-\delta\to\delta$, the Majorana states are destroyed for $\delta$ larger than a critical value of about $0.5t$, but our conclusion still holds, in that as long as the Majorana states exist, the total $S^x$ is non-zero, and $S^x$ goes to zero in the non-topological state, concomitant with the destruction of the Majoranas. This observation confirms and strengthens our conclusion that the formation of Majorana states is correlated with a total non-zero $S^x$ component of the SPDOS. More details of the  disorder analysis are presented in the SM.
We have also checked that our results do not rely on a sharp boundary between the superconducting and normal wires, thus in the SM we consider the effects of a finite SC penetration depth, and we show that our results remain generally valid. 

\section{Conclusion} We have studied the spin texture for the ABS and MBS in trivial and topological SN junctions. We have found that this is a good indicator of the topological transition. Thus, for generic SN junctions the total spin-polarized density of states $S^x$ maps out accurately the topological phase diagram, which could be measured in spin-resolved transport experiments. In a strictly 1D SN junction $S^z$ component is symmetric between positive and negative energies only in the topological phase, while exhibiting a positive-negative energy asymmetry in the trivial phase. Moreover, in this limit the $S^x$ spin accumulation at the interface changes sign at the transition. Such features are unique to the SPDOS and provide an unequivocal signature of the topological phase transition. It would be interesting to confirm our observations for more realistic calculations for junctions made using InAs and InSb wires and consider their characteristic lattice structure, size, spin-orbit coupling, NS interface properties, etc. However, as described above, we argue that our observations are a direct result of the intrinsic physics associated with the formation of Majorana states, and are thus quite generic and should hold for any quasi-1D wire. In contrast to the observation of a zero-energy peak, the features described here are extremely unlikely to have alternative explanations. Thus we claim that the transition between the topological and trivial phase could be directly and unequivocally visualized in either spin-resolved transport or spin-polarized tunneling spectroscopy experiments. 

\section*{Acknowledgments}
We would like to thank Pascal Simon, Denis Chevallier and Clement Dutreix for interesting discussions.
This work is supported by the ERC Starting Independent Researcher Grant NANOGRAPHENE 256965. Support for this research at Michigan State University (N.S.) was provided by the Institute for Mathematical and Theoretical Physics with funding from the office of the Vice President for Research and Graduate Studies.

\appendix

\section{Majorana polarization}

The Majorana polarization (MP) was recently introduced\cite{Sedlmayr2015b,Sedlmayr2016} as a theoretical tool for analyzing and testing for Majorana bound states (MBS).
A MBS is by definition both an eigenstate of the Hamiltonian, $H$, under consideration and of the particle hole operator, $\C={\bm \sigma}^y{\bm \tau}^y\hat K$. We have set the arbitrary phase of the particle-hole operator here to zero and  $\hat K$ is the complex-conjugation operator. ${\bm \sigma}$ and ${\bm \tau}$ are Pauli matrices in the spin and particle-hole spaces respectively and we work in the Nambu basis $\Psi_{j\ell}=(\psi_{\uparrow,j\ell},\psi_{\downarrow,j\ell},\psi^{\dag}_{\downarrow,j\ell},-\psi^{\dag}_{\uparrow,j\ell})^T$. The Hamiltonian anti-commutes with this operator, $\{\C,H\}=0$, and $\C^2=1$.

The local MP is the name given to a local projection of the expectation value of this operator with respect to some state. Therefore a MBS spatially localized inside a region $\mathcal{R}$ has the property that $C=1$ where
\begin{equation}\label{mpt}
C=\frac{\left|\sum_{{\vec r}\in \mathcal{R}}\langle\Psi|\C{\hat r}|\Psi\rangle\right|}{\sum_{{\vec r}\in \mathcal{R}}\langle\Psi|{\hat r}|\Psi\rangle}\,.
\end{equation}
${\hat r}_j$ is the projection onto a site $j$. The local MP is
\begin{equation}
C(j)=\langle\Psi|\C{\hat r}_j|\Psi\rangle=-2\sum_\sigma\sigma u_{j\sigma}v_{j\sigma}
\label{mp1}
\end{equation}
for a wavefunction $(u^n_{\uparrow,j\ell},u^n_{\downarrow,j\ell},v^n_{\downarrow,j\ell},v^n_{\uparrow,j\ell})$.

Note that although in general the local MP $C(j)$ is a complex quantity, as the Hamiltonians we consider here are real it is always possible to fix the phase of $\C$ such that $C(j)$ is real. This allows us to plot the local MP as a usual real observable.

\section{The effects of disorder}

Here we show some results for particular disorder realizations in the 1D SN wires considered in the main paper. The disorder is implemented in the following way: On every site of the lattice an onsite potential of a strength varying randomly from $-\delta\to\delta$ is applied. For the results in Fig.~\ref{Fig5} 100 different realizations of disorder were averaged over.

\begin{figure}
\includegraphics[width=0.9\columnwidth]{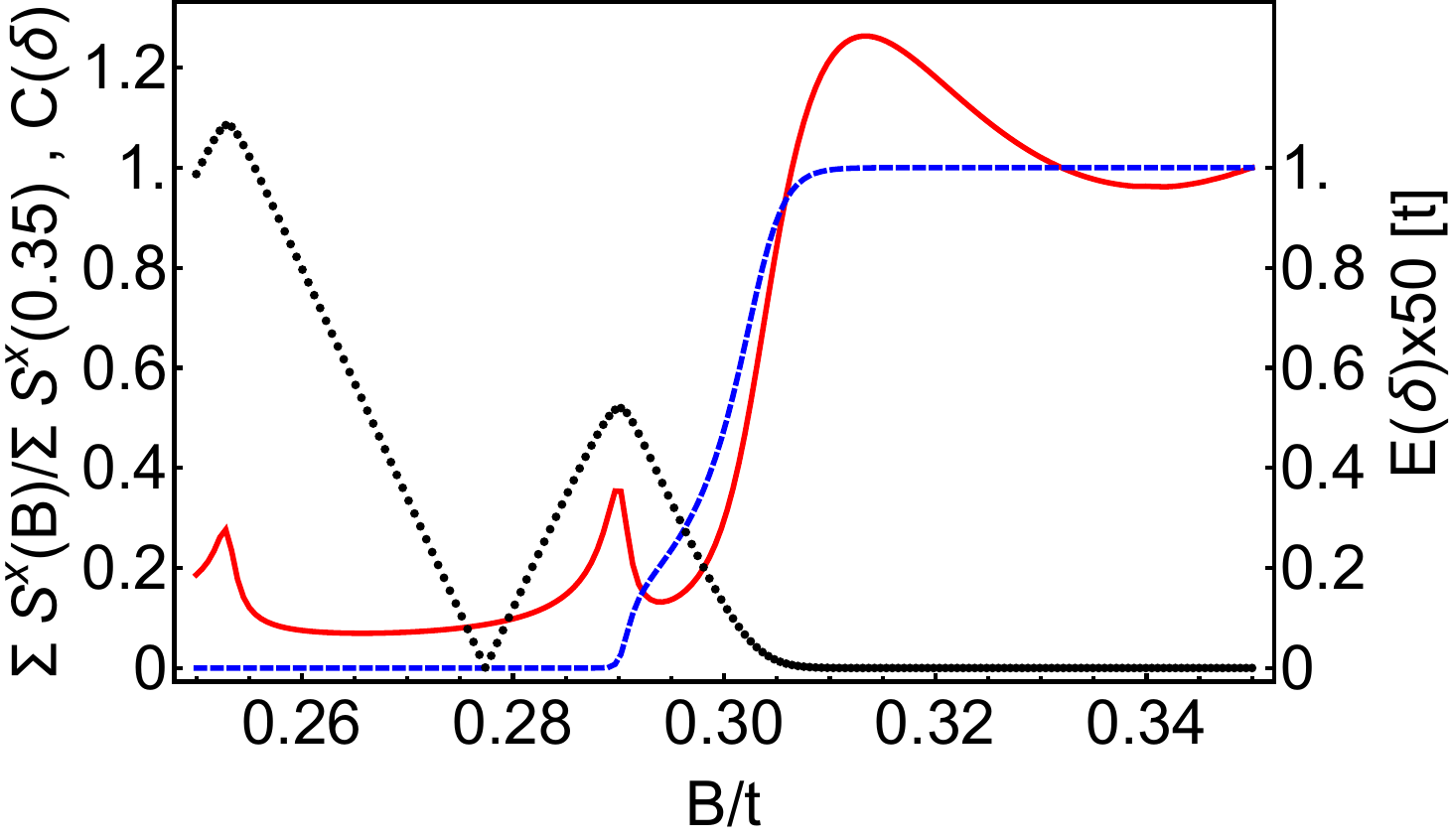}
\caption{(Color online) The total  $S^x$ spin polarization (red, solid) as a function of the magnetic field $B$, when following the evolution of the MBS into an ABS doublet for a particular disorder realization of strength $\delta=0.1t$. This is compared to the Majorana polarization $C(\delta)$ (blue, dashed) and the energy of the lowest-energy state $E(\delta)$ (black, dotted). The total spin has been scaled in units of $\sum_jS^x(B=0.35)=-0.175$.}
\label{FigSM1}
\end{figure}

In Fig.~\ref{FigSM1} we plot the total spin polarization component $S^x$ and the total Majorana polarization $C(j)$ as a function of the magnetic field for the lowest-energy state of the system for a particular disorder realization of strength $\delta=0.1t$. We note that the transition between the topological and non-topological phases is still visible in the total spin component $S^x$; this is reduced concomitantly with the drop in the total Majorana polarization and with the increase in the energy of the lowest-energy state characteristic of the transition to the non-topological phase.

\section{Finite penetration depth into the normal wire}

\begin{figure}
\includegraphics[width=0.9\columnwidth]{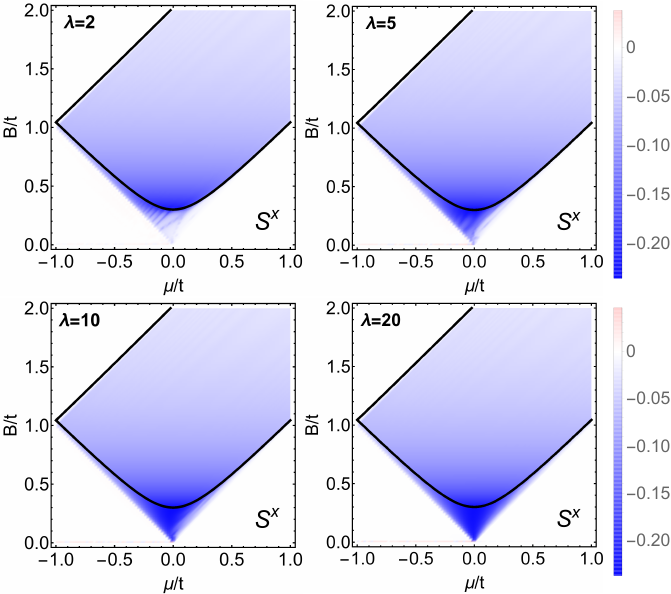}
\caption{(Color online) The total $S^x$ spin polarization in a strictly-1D wire, as a function of $\mu$ and $B$ when following the evolution of the MBS into an ABS double for different penetration depths $\lambda$. The phase diagram obtained using topological invariant calculations is very accurately recovered, except for the small region near $\mu\approx B\approx 0$. The blue regions correspond to a non-zero total SPDOS while the solid lines correspond to the analytically calculated phase boundaries. We use $\alpha=0.15t$ and $\Delta=0.3t$ in all panels. We take $N=161$.}
\label{FigSM2}
\end{figure}

\begin{figure}
\includegraphics[width=0.9\columnwidth]{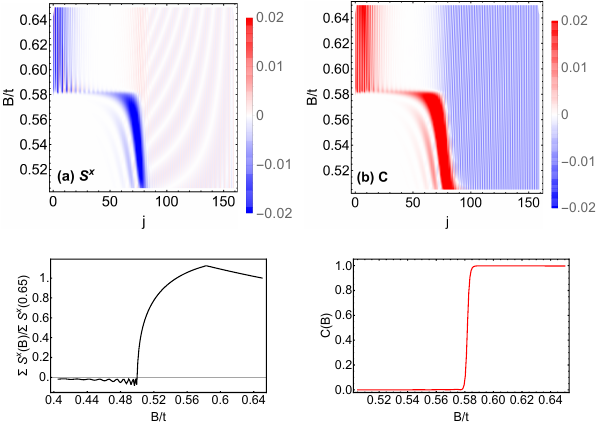}
\caption{(Color online) The $S^x$ spin polarization (panel a), and local Majorana polarization $C$ (panel b), as a function of position and $B$ when following the evolution of the MBS into an ABS doublet at $\mu=0.5t$.
The spatial variation of the superconducting pairing has a penetration depth of $\lambda=5$. To enhance contrast in panels (a) and (b) the large external MBS peaks are `cut'-out, so that their actual intensity does not show on the color scales.
The bottom panels show the total $S^x$ summed over the whole system, and C, see Eq.~\eqref{mpt}, as a function of the magnetic field for $\lambda=5$. The total spin has been scaled compared to $\sum_jS^x(B=0.65)=-0.106$.}
\label{FigSM3}
\end{figure}

We now turn to what happens if the boundary between the superconducting wire and the normal wire is not sharp. We model the change in the proximity induced s-wave pairing in this case as $\Delta_i=(1-\tanh[(i-(N+1)/2)/\lambda])/2$, with $\lambda$ describing the sharpness of the boundary. $N$ is the total length of the SN system and $(N+1)/2$ the location of the boundary between S and N. 
As can be seen in Fig.~\ref{FigSM2} the total spin criterion for the topological phase transition remains valid even for large penetration depths $\lambda$, only failing near $\mu\approx 0$. The smoothness of the boundary introduces an additional energy scale $E_\lambda\sim\lambda^{-1}$, and our results hold even near $\mu=0$ provided the ABS are below this energy scale. Away from $\mu\approx 0$ the results always hold. As an example, in Fig.~\ref{FigSM3} we show the spin polarization and Majorana polarization across the phase transition at $\mu=0.5$. For even relatively long penetration depths we see the same physics, i.e. a non-zero component of the total $S^x$ in the topological phase decreasing to zero in the non-topological phase.

\section{Impurity at the boundary}

\begin{figure}
\includegraphics[width=0.9\columnwidth]{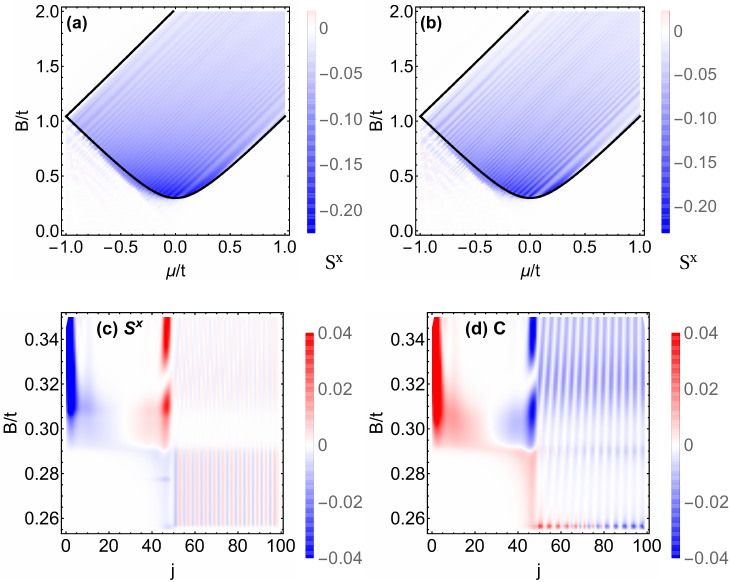}
\caption{(Color online) The total $S^x$ spin polarization in a strictly-1D wire, as a function of $\mu$ and $B$ when following the evolution of the MBS into an ABS double for different impurity strengths at the boundary: (a) $\gamma=0.5t$, and (b) $\gamma=t$. The phase diagram obtained using topological invariant calculations is very accurately recovered: the blue regions correspond to a non-zero total SPDOS while the solid lines correspond to the analytically calculated phase boundaries. We use $\alpha=0.15t$ and $\Delta=0.3t$ in all panels. For (a,b) we used $N=161$. The bottom two panels show the spin polarization $S^x$ (panel c), and local Majorana polarization $C$ (panel d), as a function of position and $B$ when following the evolution of the MBS into an ABS doublet at $\mu=0$ and $\gamma=0.5t$.  To enhance contrast in panels (c) and (d) the large external MBS peaks are `cut'-out, so that their actual intensity does not show on the color scales.}
\label{FigSM4}
\end{figure}

Fig.~\ref{FigSM4} shows the effect of including a local impurity term of strength $\gamma$ at the SN boundary. This is included in the Hamiltonian as a term $\gamma\Psi^\dagger_{L_1}\Psi_{L_1}$. Even for a large local impurity $\gamma\sim t$ the phase diagram is still accurately recovered. Additionally the local behaviour of the spin density remains unchanged by impurities of this strength.

\section{Experimental feasability}

The signatures of the topological phase transition discussed above could be experimentally verified in several ways. The local spin build up seen in the stricly 1D wires, see Fig.~\ref{Fig1}(b), would be visible in an energy and spin resolved STM experiment focused at the SN boundary. More generally the total spin build up, can be measured by a spin-resolved transport measurement, see Fig.~\ref{Fig1}(b). The two leads are held at a potential of $\pm eV/2$ respectively. If the top lead injects a spin polarized current parallel to the $S^x$ direction in the wire then from Fermi's golden rule we find the linear response differential conductance:
\begin{equation}
G^x_\pm(B)\equiv\frac{G^x_{\uparrow,\downarrow}(B)}{G^x_0}=\rho(\omega=0,B)\pm S^x(\omega=0,B)\,,
\end{equation}
where $G^x_0=2\pi e^2|T|^2/\hbar$ is the conductance quantum multiplied by the coupling between the leads and SN wire, $|T|^2$.
From these we have direct qualitative access to the results of Fig.~\ref{Fig2} demonstrating the topological phase transition.

\end{document}